\documentclass[smallextended]{svjour3}
\usepackage{amsmath,amssymb}
\usepackage{bm}
\RequirePackage{graphicx}
\RequirePackage{mathptmx}
\RequirePackage{latexsym}
\RequirePackage[numbers,sort&compress]{natbib}
\RequirePackage[colorlinks,citecolor=blue,urlcolor=blue,linkcolor=blue]{hyperref}

\begin{document}

\title{A minimal fractional deformation of Newtonian gravity: a unified description of cosmology and weak-field tests}

\author{S. M. M. Rasouli}

\institute{
S. M. M. Rasouli \at
Departamento de F\'{\i}sica,\\
Centro de Matem\'{a}tica e Aplica\c{c}\~oes (CMA-UBI),\\
Universidade da Beira Interior,\\
Rua Marqu\^es d'\'Avila e Bolama,\\
6200-001 Covilh\~a, Portugal\\
\email{mrasouli@ubi.pt}
}

\authorrunning{S. M. M. Rasouli}
\date{\today}
\maketitle

\begin{abstract}

We investigate a minimal fractional deformation of
Newtonian gravity as a unified effective framework
for cosmology and weak-field gravitational tests.
 Here we demonstrate that a single-parameter fractional extension of the Newtonian action, motivated by fractal frameworks in field theory with a Stieltjes-type measure, gives rise to an effective framework capable of reproducing gravitational dynamics across vastly different scales. 
Previous studies of this framework have addressed cosmic
evolution, including a nonsingular pre-inflationary phase,
an inflationary attractor, radiation- and matter-dominated
epochs, and the present accelerated expansion.
The evolution of matter density perturbations has also
been investigated, with results compatible with
cosmological constraints.
These applications share a unified effective potential
description without introducing additional dynamical
degrees of freedom.
In this paper, we extend this construction to weak-field gravity, showing that key 
tests such as the perihelion precession of Mercury and the deflection of 
light grazing the solar limb are consistently described within the same potential, with all 
constraints pointing to only a small deviation from standard Newtonian dynamics.
These results support a unified effective description of
cosmological evolution and weak-field gravity, providing
a basis for exploring the smallness of the effective
cosmological constant, the hierarchy of Hubble scales,
and the Hubble tension.

\end{abstract}

\keywords{modified gravity \and Stieltjes measure \and fractional dynamics \and Newtonian gravity \and $\Lambda$CDM model \and weak field tests \and Hubble tension \and  hierarchy of cosmological scales
}

\maketitle
\section{Introduction}

The description of gravity across widely separated physical scales remains a central question in theoretical physics. General relativity (GR) successfully describes gravitational phenomena from Solar-System tests to cosmology~\cite{will2014confrontation}. Whether key features of relativistic gravitational dynamics can emerge from a simpler underlying framework nevertheless remains an important question. In particular, can a minimal modification of Newtonian dynamics provide a common effective description of cosmological evolution and local gravitational phenomena?

In practice, explaining the full cosmic evolution within GR typically requires introducing additional ingredients, such as dark energy, scalar fields, or modifications of gravity \cite{capozziello2010beyond}. Moreover, several fundamental cosmological puzzles remain unresolved, including the cosmological constant problem \cite{weinberg1989cosmological}, the hierarchy between the inflationary and present Hubble scales \cite{lyth1999particle}, and the current tension in the measurements of the Hubble constant \cite{Planck:2018vyg,verde2019tensions,riess2022comprehensive}.

In this direction, it is particularly compelling to explore whether a minimal extension of Newtonian gravity could provide an effective framework capable of reproducing both cosmological evolution and local gravitational phenomena, without introducing additional degrees of freedom. Such a possibility would offer a unified description of gravitational dynamics across scales within a conceptually simpler setting, and could provide new insight into the origin of relativistic behavior in gravity.

As a minimal realization of such an interesting  framework, a fractional extension of the Newtonian action was 
introduced in Ref.~\cite{rasouli2026emergent},
inspired by fractal spacetime frameworks in field theory~\cite{calcagni2010fractal,calcagni2010quantum}.
The construction is characterized by a single deformation parameter $\alpha$:
\begin{eqnarray}\label{New-fr-action}
 S_{\alpha}=
 \!\!\frac{1}{\Gamma(\alpha)}\int_{0}^{\bar t} \xi({\tau} )\left(T-V_{\rm eff}\right)\,d{\tau},
\end{eqnarray}
where $\xi({\tau})\equiv[({\bar t} -\tau)/{\tau_0}]^{\alpha-1}$ is a time-dependent kernel; $T\equiv1/2 \,\,m (d\mathbf{r}/d\tau)^2$ is the standard kinetic energy and $V_{\rm eff}=V_{\rm eff}(\mathbf{r},\alpha)$ is an effective potential energy. In the special case $\alpha=1$, the kernel becomes unity. With $V_{\rm eff}(\mathbf{r}, 1)=V(\mathbf{r})$, the action \eqref{New-fr-action} reduces to that of standard Newtonian mechanics.

The fractional action \eqref{New-fr-action} can be interpreted, at an effective level, as being motivated by frameworks of fractional geometry in field theory, where a generalized spacetime measure, such as a Stieltjes-type measure, naturally gives rise to fractional integrals and memory kernels. In this perspective, the fractional parameter may be associated with an effective (Hausdorff) dimension of spacetime \cite{calcagni2010quantum,calcagni2010fractal}.
This interpretation provides a natural link between the fractional deformation and an effective modification of spacetime structure.

The presence of a time-dependent kernel $\xi(\tau)$ in the action introduces an additional friction-like term in the equations of motion. This modification leads to a generalized contribution to the total energy, naturally interpreted as a fractional kinetic term corresponding to a logarithmic-time average of the standard kinetic energy.
Within this framework, it was shown that a
relativistic cosmology can dynamically emerge from the modified
Newtonian action \cite{rasouli2026emergent}. 

The resulting emergent cosmological model reproduces the full sequence of cosmic evolution, from a nonsingular pre-inflationary phase and an inflationary attractor to the radiation- and matter-dominated eras and the present accelerated expansion, while remaining consistent with the current observational data \cite{rasouli2026emergent,rasouli2026inflation}.
The growth of large-scale structure is also described 
within the same framework through a fractional 
generalization of the standard perturbation equation \cite{rasouli2026evolution}.

Both the cosmological background evolution and the growth
of matter perturbations are described within a unified
framework based on a single effective gravitational
potential and a single deformation parameter $\alpha$.
Within this framework, the observed cosmological dynamics are reproduced without invoking exotic matter components or a cosmological constant.
The previous cosmological analyses favor values of
$\alpha$ close to unity~\cite{rasouli2026emergent,rasouli2026inflation,rasouli2026evolution}.
To examine how small departures from Newtonian dynamics
affect weak-field observables, we focus here on the regime
\begin{equation}\label{alpha-cons}
|\alpha-1|\equiv\varepsilon\ll1  .
\end{equation}
This regime is also compatible with the small departures
from unity considered in our earlier analysis of a
prolonged radiation-dominated era~\cite{rasouli2026inflation}.

Motivated by these results, we extend the minimal fractional framework to the weak-field regime and investigate whether it can consistently describe key Solar-System tests of gravity, including the perihelion precession of Mercury (PPM) and the gravitational deflection of light (GDL), in agreement with general relativity.
Combined with our previous cosmological results \cite{rasouli2026emergent,rasouli2026inflation,rasouli2026evolution}, this allows us to assess whether the same minimal construction captures the essential features expected from a viable gravitational theory across a wide range of physical scales.
Furthermore, we explore whether this unified framework can provide new insight into longstanding problems of modern cosmology, including the cosmological constant problem, the hierarchy between the inflationary and present Hubble scales, and the current tension in measurements of the Hubble constant.



\section{Fractional Binet's equation and the PPM}

To test the fractional framework in the weak-field regime, we consider the motion of a test particle in a central effective potential $V_{\rm eff}=V_{\rm eff}(r;\alpha)$. Adopting planar polar coordinates $(r,\phi)$ and using \eqref{New-fr-action}, the modified angular momentum $L_\phi$ takes the form
\begin{equation}\label{cons-of-motion}
h_{\rm frac}\equiv \frac{L_\phi}{m} 
=\xi(t)\, r^2\dot\phi 
={\rm constant},
\end{equation}
where we set ${\bar t} - \tau\equiv t$ and a dot denotes the derivative with respect to $t$.

Eliminating $\dot\phi$ in favor of the conserved quantity $L_\phi$, we construct the Routhian \cite{routh1877elementary,safko2002classical}
\begin{equation}\label{Routhian-def}
\mathcal{R}(r,\dot r;p_\phi,t)\equiv \xi(t)\,\mathcal{L}(r,\dot r,\dot\phi)-L_\phi\dot\phi,
\end{equation}
which leads to the radial equation:
\begin{equation}\label{master-r}
\ddot r+\frac{\dot\xi}{\xi}\,\dot r
-\frac{L_\phi^2}{m^2\xi^2}\,\frac{1}{r^{3}}
+\frac{1}{m}\frac{dV_{\rm eff}}{dr}=0.
\end{equation}

Using the standard variable $u(\phi)=1/r(\phi)$, the equation \eqref{master-r} can be recast as
\begin{equation}\label{master-u}
u''+u=
\frac{\,\xi^2}{m h_{\rm frac}^2 u^2}
\frac{dV_{\rm eff}(r;\alpha)}{dr}\Bigg\lvert_{r=1/u}\,,
\end{equation}
where a prime denotes the differentiation with respect to the polar coordinate $\phi$.
Interestingly, the explicit terms proportional to $\dot{\xi}$
cancel identically in transforming the radial equation to the
orbital variable $u(\phi)$, while the fractional time dependence
enters through the multiplicative factor $\xi^2(t)$.

Equation~\eqref{master-u} constitutes the fractional generalization of the Binet's equation for a general central potential. Within this framework, the parameter $\alpha$ controls the deviation from Newtonian dynamics, while the limit $\alpha = 1$ recovers the standard Keplerian motion \cite{thorne2000gravitation,d2023introducing}.

In the weak--deformation regime \eqref{alpha-cons}, the time-dependent kernel admits the expansion  
\begin{equation}\label{xi-cons}
\xi(t)=1\pm\varepsilon\ln\!\left(\frac{t}{t_*}\right)
+\mathcal{O}\!\left(\varepsilon^2\ln^2\!\frac{t}{t_*}\right),
\end{equation}
where the upper and lower
signs correspond to $\alpha>1$ and $\alpha<1$, respectively.
This expansion requires $\epsilon|\ln(t/t_*)|\ll1$.

To test the model 
against the Solar-System observables (PPM and GDL), we adopt a unified effective potential $
V_{\rm eff}(r;\alpha)=V_{N}(r)+V_{\alpha}(r;\alpha)$, 
where \(V_N(r)=-\mu/r\) is the Newtonian contribution and \(V_\alpha\) represents subleading corrections induced by the fractional deformation \cite{rasouli2026emergent,rasouli2026inflation}.  From symmetry and dimensional considerations, the most general weak-field potential compatible with spherical symmetry can be written as
\begin{equation}\label{Veff-common}
V_{\rm eff}(r;\alpha)
=
-\frac{m \mu}{r}-\frac{m \mu\, A(\alpha)}{r^3}-
\frac{m \mu B(\alpha)}{r}\, e^{-r/\lambda},
\end{equation}
where $m$ is the reduced mass and $\mu\equiv G(m_1+m_2)$. In the test-mass limit, $m_2\ll m_1$, assuming $m_1\equiv M$, we get $\mu\simeq GM$ . 
The scale $\lambda={\rm constant}$ characterizes the Yukawa interaction, while $A(\alpha)$ and $B(\alpha)$ set 
the strength of the  fractional corrections that vanish as $\alpha=1$, ensuring the recovery of the standard Newtonian interaction.

To facilitate comparison, the relativistic Binet's equation in general relativity reads \cite{d2023introducing}
\begin{equation}\label{GR-binet}
    u''+u=\frac{m_{\rm s}}{h^2}+3m_{\rm s}u^2,
\end{equation}
where $h=r^2\dot{\phi}$ is the conserved specific angular momentum, and the last term represents the leading post-Newtonian correction responsible for the perihelion precession. 

To establish correspondence with this limit, we parametrize the fractional correction as $A(\alpha)=\ell^2 f(\alpha)$, where $\ell$ denotes a characteristic length scale of the fractional deformation, and \(f(\alpha)\) is a dimensionless 
function where $f(1)=0$, ensuring the recovery of the Newtonian regime. Expanding around this limit, we write $f(\alpha)\simeq c_1(\alpha-1)+\mathcal{O}((\alpha-1)^2)$ .
At leading order, this implies
\begin{equation}\label{A-alpha}
A(\alpha)\simeq |\alpha-1|\ell^2.
\end{equation}
Thus, in the weak-deformation regime \(|\alpha - 1| \ll 1\), a broad class of functions satisfying \(f(1)=0\) effectively reduces to the same linear behavior, making the results largely insensitive to the specific choice of \(f(\alpha)\). A similar expansion can be introduced for the second coefficient $B(\alpha)$, consistent with its physical dimension.


In the weak-deformation regime \eqref{alpha-cons}, the coefficient 
$A(\alpha)=|\alpha-1|\ell^2$ remains small unless the characteristic scale $\ell$ is sufficiently large. This suggests interpreting $\ell$ as a macroscopic length scale controlling the strength of the fractional correction. 
A similar argument applies to $B(\alpha)$. Since it multiplies the Yukawa contribution, a physically consistent deformation should not suppress this term unnaturally, implying $B(\alpha)$ is generically of order unity in the absence of additional hierarchies. Consequently, observable deviations from Newtonian gravity are primarily governed by the scale $\ell$ and the deformation parameter $|\alpha-1|$.

For planetary motion, the Yukawa contribution is exponentially suppressed, scaling as $e^{-r/\lambda}$. In the regime $r \gg \lambda$ its contribution rapidly becomes negligible. 
For Mercury, with an orbital radius $r_M \simeq 5.8\times10^{10}\,{\rm m}$, and a characteristic solar scale $\lambda \sim R_\odot \simeq 7\times10^{8}\,{\rm m}$, the Yukawa term is effectively inactive.
As a result, the dominant deviation from Newtonian dynamics arises from the power-law correction proportional to $r^{-3}$. Consequently, Solar-System tests such as the perihelion precession of Mercury provide a direct probe of the coefficient $A(\alpha)$.

Therefore, neglecting the Yukawa contribution and substituting \eqref{Veff-common} and \eqref{A-alpha} into the fractional Binet's equation \eqref{master-u}, we obtain
\begin{equation}\label{Binet-Frac-1}
u''+u=\frac{\xi^2(t) GM}{h_{\rm frac}^2}\Big[1+3|\alpha-1|\,\ell^2u^2\Big].
\end{equation}

In the weak-deformation regime  \eqref{alpha-cons}, the logarithmic dependence in $\xi^2(t)$ provides only a sub-leading correction over Solar--System timescales, and can therefore be treated as effectively constant over one orbital period: $\xi^2(t)\approx {\bar{\xi}}^2$ .  Under this approximation, equation \eqref{Binet-Frac-1} reduces to the same form as the relativistic Binet's equation \eqref{GR-binet}, with the quadratic term acting as a small perturbation to the Keplerian orbit. The solution can then be obtained perturbatively by expanding $
u = u_0 + u_1$, with $u_1\ll u_0$ \cite{d2023introducing}: 
\begin{equation}
u(\phi)\simeq
C_{\rm frac}\Big[1+e\cos((1-\sigma)\phi)\Big],
\end{equation}
where $e$ is the orbital eccentricity and 
\begin{equation}
\sigma \equiv 3|\alpha-1|\,\ell^2C_{\rm frac}^2, \qquad
C_{\rm frac}\equiv{\bar{\xi}^2 GM}/{h_{\rm frac}^2}\,.
\end{equation}
The orbit is therefore no longer exactly periodic with period $2\pi$, but instead has angular period $\phi_{\rm period} = 2\pi/(1-\sigma)$. This directly yields the perihelion shift per orbit as
\begin{equation}
\Delta\phi_{\rm frac}
=
6\pi |\alpha-1|\,\ell^2
\left(
\frac{\bar{\xi}^2 GM}{h_{\rm frac}^2}
\right)^2.
\end{equation}
In the regime $\varepsilon\ll1$, equations \eqref{cons-of-motion} and~\eqref{xi-cons} imply that, to leading order, the fractional angular momentum deviates only slightly from the standard counterpart, so that $h_{\rm frac}^2\simeq h^2$, and we may consider $\bar{\xi}^2\simeq1$ and $C_{\rm frac}=C={ GM}/{h ^2}$ \cite{d2023introducing}.
Therefore, the ratio of the fractional prediction to the relativistic one becomes
\begin{equation}
\frac{\Delta\phi_{\rm frac}}{\Delta\phi_{\rm GR}}
=
\frac{|\alpha-1|\,\ell^2}{h^2}.
\end{equation}

Therefore, consistency with the observations requires 
$|\alpha-1|\simeq {h^2}/{\ell^2}$.
This shows that Solar-System tests, and in particular perihelion precession measurements, provide a direct observational constraint on the fractional deformation parameter $\alpha$, once the characteristic length scale $\ell$ is fixed.

 As discussed earlier, $\ell$ should correspond to a macroscopic length scale associated with the system. A natural choice is $
\ell=\sqrt{aR_\odot}$, corresponding to the geometric mean of the orbital radius $a$ and the radius of the central body.  
For the Sun, one has $m_{s}=GM_\odot\approx1.47\times10^3\ \mathrm{m}$,
and $
R_\odot\approx7\times10^8\ \mathrm{m}$.  
For Mercury, the semi-major 
axis of the orbit is $
a \approx 5.79\times10^{10}\ \mathrm{m}$, yielding $\ell\approx6\times10^9\mathrm{m}$.
Using $h^2=m_{s} a(1-e^2) $, where $e=0.2056$ is the orbital eccentricity of Mercury, the observational bound from the perihelion shift gives
\begin{equation}
|\alpha-1|\simeq\frac{h^2}{\ell^2}
           \simeq\frac{m_{\rm s}a}{aR_\odot}
           =\frac{m_{\rm s}}{R_\odot}\simeq2\times10^{-6},
\end{equation}
This result is fully consistent with the weak-deformation assumption \eqref{alpha-cons}.

\section{Light Deflection in the Fractional Newtonian Framework}

To study the gravitational deflection of light (GDL) within the FNF, we adopt the same effective potential \eqref{Veff-common} employed in the orbital analysis.
 This unified treatment, based on a single potential for both massive and massless probes, provides a consistent description of weak-field gravitational phenomena.

To describe light propagation, we adopt the phenomenological
weak-field refractive index $n \simeq 1-\Phi$, where
$\Phi \equiv V_{\mathrm{eff}}/m$ and we use units $c=1$.
The deflection angle of a photon passing at impact parameter $b$ is then given by
\begin{equation}\label{fermat-eq}
\delta
=
-\int_{-\infty}^{+\infty}
\frac{\partial \Phi}{\partial b}\,dz,
\end{equation}
where $r=\sqrt{b^2+z^2}$. Equation \eqref{fermat-eq} follows from Fermat's principle, to leading
order in the weak-field and small-deflection approximations,
using the adopted refractive index.

For the Newtonian contribution, $\Phi_N=-GM/r$,
equation~\eqref{fermat-eq} gives
$|\delta_N|=2GM/b$. The negative sign of the signed
deflection indicates bending toward the source.
The magnitude is one half of the general-relativistic result.

For the power-law contribution,
$\Phi_A=-GM A(\alpha)/r^3$, the same expression yields
\begin{equation}\label{deltaA_deltaN}
|\delta_A|=\frac{8GM A(\alpha)}{b^3},
\qquad
\frac{|\delta_A|}{|\delta_N|}
=\frac{4A(\alpha)}{b^2},
\end{equation}
where $A(\alpha)>0$ in the adopted parametrization.
Using the illustrative value
$A(\alpha)\simeq8\times10^{13}\,\mathrm{m}^2$
from the Mercury analysis and
$b\simeq R_\odot\simeq7\times10^8\,\mathrm{m}$,
this ratio is approximately $6.5\times10^{-4}$.
We therefore neglect this contribution in the following
leading-order estimate.

For the Yukawa contribution,
$\Phi_Y=-GM B(\alpha)e^{-r/\lambda}/r$,
evaluating equation~\eqref{fermat-eq} along the
unperturbed straight-line trajectory gives
\begin{equation}
|\delta_Y|
=\frac{2GM|B(\alpha)|}{\lambda}
K_1\!\left(\frac{b}{\lambda}\right),
\end{equation}
where $K_1$ denotes the modified Bessel function
of the second kind of order one.

Neglecting the power-law contribution in the following
leading-order estimate and taking $B(\alpha)>0$,
the total deflection angle becomes
\begin{equation}\label{Tot-Def}
|\delta|\simeq\frac{2GM}{b}
\left[
1+B(\alpha)\frac{b}{\lambda}
K_1\!\left(\frac{b}{\lambda}\right)
\right].
\end{equation}

Matching this expression to the general-relativistic
prediction $\delta_{\mathrm{GR}}=4GM/b$ at the chosen
impact parameter requires
\begin{equation}\label{cons-GR}
1+B(\alpha)\frac{b}{\lambda}
K_1\!\left(\frac{b}{\lambda}\right)
\simeq 2.
\end{equation}

For light rays grazing the Sun, the impact parameter
is approximately the solar radius, $b\simeq R_\odot$.
We take $\lambda=R_\odot$ as a representative
source-scale choice, giving $b/\lambda\simeq1$.
With this interaction range, the Yukawa correction
can contribute to light bending near the solar limb,
while its exponential decay suppresses its influence
at planetary distances.

Using $K_1(1)\simeq0.6019$, equation~\eqref{cons-GR} gives
\[
B(\alpha)\simeq\frac{1}{K_1(1)}\simeq1.66.
\]
Thus, the required Yukawa amplitude remains of order unity.
To relate this amplitude to the fractional parameter, we adopt
the phenomenological parametrization
$B(\alpha)=(\alpha-1)/\epsilon_g$, where
$\epsilon_g=m_s/R_\odot$ and $m_s=GM_\odot$ in units $c=1$.
With $m_s\simeq1.47\times10^3\,\mathrm{m}$ and
$R_\odot\simeq7\times10^8\,\mathrm{m}$, we obtain
\[
\alpha-1
\simeq B(\alpha)\epsilon_g
\simeq1.66\times2.1\times10^{-6}
\simeq3.5\times10^{-6}.
\]
The attractive Yukawa branch considered here therefore
corresponds to $\alpha>1$.

The light-deflection estimate and the Mercury estimate
both place $|\alpha-1|$ at the order of $10^{-6}$ under
their respective scale choices. Their numerical values
are not identical. They should therefore be interpreted
as illustrative matching estimates within the adopted
effective description, rather than as a joint
observational constraint.


In summary, the two classical weak-field tests yield
illustrative estimates of the fractional parameter within
the adopted effective description. With the respective
scale choices and parametrizations used here, the Mercury
analysis gives $|\alpha-1|\simeq2\times10^{-6}$, while
the light-deflection analysis gives
$\alpha-1\simeq3.5\times10^{-6}$.
Both estimates are therefore of order $10^{-6}$,
although they do not constitute a joint observational
constraint.

These near-Newtonian values can also be compared with
the cosmological parameter ranges previously studied
within the fractional framework. In particular,
$|\alpha-1|<0.2$ was obtained for the present
Universe~\cite{rasouli2026emergent}, while
near-Newtonian values were also considered in the
early-universe analysis~\cite{rasouli2026inflation}.
The matter-perturbation analysis gives the upper bound
$\alpha\lesssim1.07$~\cite{rasouli2026evolution}.
The illustrative weak-field estimates lie within
these broad cosmological bounds. This comparison
indicates compatibility at the level of the quoted
parameter ranges; a joint quantitative assessment
across these regimes remains to be performed.

\section*{Summary and Outlook}

In our previous work \cite{rasouli2026emergent}, we proposed
an effective modified Newtonian action motivated by
fractional geometry frameworks with a generalized
Stieltjes-type spacetime measure
\cite{calcagni2010fractal,calcagni2010quantum}.
We have shown that an effective relativistic
Friedmann--Lemaître--Robertson--Walker cosmology can
emerge from a minimal fractional deformation of
Newtonian gravity \cite{rasouli2026emergent}.
Previous studies have investigated both background
evolution and matter density perturbations within
this framework
\cite{rasouli2026emergent,rasouli2026inflation,rasouli2026evolution}.

In the present work, we demonstrate that this minimal
setup provides a unified description of cosmological
evolution and local gravitational tests within the
same fractional framework.
The deformation parameter $\alpha$ introduces departures
from standard Newtonian dynamics that can be quantified
and constrained by observations.
These results provide a basis for revisiting questions
in relativistic cosmology within an effective description
derived from a modified Newtonian action.

For values of the fractional parameter close to unity,
the model reproduces the full sequence of cosmic evolution
within a unified framework. In particular, it accounts for
a nonsingular pre-inflationary phase, a quasi-de Sitter
inflationary epoch~\cite{rasouli2026inflation}, the subsequent radiation- and
matter-dominated eras, and the present phase of accelerated
expansion~\cite{rasouli2026emergent}. These distinct cosmological stages arise from
the same underlying fractional structure, reflecting
the scale-dependent interplay between the temporal
(kinetic) and spatial (potential) sectors of the theory.

At the level of first-order perturbations, the growth of density fluctuations is governed by a fractional generalization of the standard growth equation, leading to large-scale structure formation compatible with observational constraints under the same bound on the deformation parameter $\alpha$ \cite{rasouli2026evolution}.

An important feature of the model is that, in the special
case $\alpha=1$, the fractional action reduces to its
standard Newtonian counterpart, recovering Newtonian
gravity. In this sense, the fractional deformation
parameter $\alpha$ provides a controlled and physically
transparent realization of a generalized correspondence
principle, interpolating continuously between standard
Newtonian dynamics and its fractional extension.

Another key conceptual implication of our fractional
framework is the interplay between its temporal (kinetic)
and spatial (potential) sectors, both governed by the same
deformation parameter $\alpha$. Their relative importance
depends on the characteristic energy or time scale of
the system.
In high-energy regimes, such as the early Universe,
the rapid evolution and short characteristic time scales
allow the kernel-induced kinetic contribution to dominate,
making the fractional kinetic sector the primary driver
of the dynamics during the pre-inflationary and inflationary
phases \cite{rasouli2026inflation}.
By contrast, in low-energy, weak-field settings, such as
the Solar System, the evolution is slow and the kernel
is effectively constant over observational time scales.
The kernel-induced kinetic correction is therefore
suppressed, and the leading deviations from standard
Newtonian dynamics arise from the fractional potential
sector.
This scale-dependent behavior provides a unified physical
interpretation of the model: early-universe cosmology
predominantly probes the temporal sector, whereas local
gravitational tests probe the spatial sector. Different
physical regimes thus probe complementary aspects of
the same underlying fractional structure, suggesting
a common description of gravitational dynamics across
widely separated scales.

 
Beyond the cosmological results of our previous studies,
the fractional framework offers a perspective on two open
questions in cosmology: the smallness of the effective
cosmological constant and the large separation between
the inflationary and present-day Hubble scales.

As shown in Refs.~\cite{rasouli2026inflation,rasouli2026emergent},
the two scales are set by different functions of the same
deformation parameter $\alpha$:
\begin{eqnarray}\label{H-inf}
H_{\rm inf}(t)&\simeq&
     \sqrt{\mathcal{F}_0(\alpha)}-\frac{\alpha-1}{2\, t},\\ 
\label{H-late-time-part}
H_{\rm late}(t) &\simeq&\kappa(\alpha) - \frac{\alpha-1}{2t} + \mathcal{O}\!\left(\frac{1}{t^2}\right),
  \qquad
 \kappa t\gg 1,
\end{eqnarray}
with both functions satisfying $\mathcal{F}_0(1)=0=\kappa (1)$. 

In the present epoch, the time-dependent correction is negligible, leading to $H_0 \approx \kappa(\alpha)$ and an effective cosmological constant $\Lambda_{\mathrm{eff}} \sim 3\kappa^2(\alpha)$ that dynamically emerges from the fractional structure.
 From this perspective, the small observed value of the cosmological constant may naturally be associated with the small deviation of the fractional parameter from unity rather than requiring a finely tuned vacuum energy density.

To compare the characteristic Hubble scales, we use the
inflationary plateau and the late-time asymptotic value,
neglecting the time-dependent correction
where it is subdominant. The resulting ratio is
${H_{\rm inf}}/{H_0}\simeq{\sqrt{\mathcal F_0(\alpha)}}/{\kappa(\alpha)}$.
The hierarchy depends on
how $\sqrt{\mathcal{F}_0(\alpha)}$ and $\kappa(\alpha)$
vary near $\alpha=1$.
For an illustrative inflationary scale $H_{\mathrm{inf}} \sim 10^{37}\,\mathrm{s}^{-1}$ and the 
present Hubble scale $H_0 \sim 10^{-18}\,\mathrm{s}^{-1}$, the ratio is of order $10^{55}$.

To illustrate
the role of this functional freedom, we use $\varepsilon\equiv|\alpha-1|$
and consider the following phenomenological choice 
\[
\mathcal{F}_0(\alpha)=A\,\varepsilon,\qquad
\Lambda_{\mathrm{eff}}(\alpha)\,
=|\lambda_1|\,\varepsilon\exp(-2C_1/\varepsilon),
\]
where $A,C_1>0$  and $\lambda_1\neq0$ are constants. Both functions are set to zero at $\varepsilon=0$.
The inflationary plateau scale and the
asymptotic late-time Hubble rate then 
share the same $\sqrt{\varepsilon}$ dependence, with an additional exponential suppression in the late-time rate.
In units where $c=1$, defining $M_{\mathrm{inf}}\equiv\sqrt{A}$ and
$M_{\mathrm{late}}\equiv\sqrt{|\lambda_1|/3}$, we obtain the ratio
of the inflationary plateau scale to the asymptotic
late-time Hubble rate:
\[
\frac{H_{\mathrm{inf}}}{\kappa}
\simeq
\frac{M_{\mathrm{inf}}}{M_{\mathrm{late}}}
\exp(C_1/\varepsilon).
\]
For illustration, taking $M_{\mathrm{inf}}=M_{\mathrm{late}}$,
$\varepsilon=0.10$, and $C_1\simeq12.6642$ gives a ratio of order
$10^{55}$. In this example, the hierarchy arises from the
exponential suppression of the late-time rate, even though
the two prefactors are equal. The value of $C_1$ is chosen
to illustrate the target hierarchy. Identifying this ratio
with $H_{\mathrm{inf}}/H_0$ further requires the asymptotic
approximation $H_0\simeq\kappa$ to hold at the present epoch.
Restoring $c$, for the phenomenological choice
$\kappa=M_{\mathrm{late}}\sqrt{\varepsilon}\exp(-C_1/\varepsilon)$,
the effective cosmological constant follows directly from
the asymptotic late-time Hubble rate:
\begin{equation}
\Lambda_{\mathrm{eff}}
= \frac{3\kappa^2}{c^2}
= \frac{3M_{\mathrm{late}}^2}{c^2}\,
  \varepsilon\exp\!\left(-\frac{2C_1}{\varepsilon}\right).
\end{equation}
For $\varepsilon=0.10$ and $C_1\simeq12.6642$, the dimensionless
factor $\varepsilon\exp(-2C_1/\varepsilon)$ is of order $10^{-111}$.
To fix the overall scale, we take the inflationary plateau
rate $H_{\mathrm{inf}}\sim10^{37}\,\mathrm{s}^{-1}$ as an
illustrative input. With equal prefactors and
$H_{\mathrm{inf}}\simeq M_{\mathrm{inf}}\sqrt{\varepsilon}$,
this gives
\[
M_{\mathrm{late}}=M_{\mathrm{inf}}
\simeq \frac{H_{\mathrm{inf}}}{\sqrt{\varepsilon}}
\sim 3.16\times10^{37}\,\mathrm{s}^{-1}.
\]
Substituting this value into the expression above yields
$\Lambda_{\mathrm{eff}}\sim3.3\times10^{-53}\,\mathrm{m}^{-2}$,
which is of the same order of magnitude as the value inferred in flat
$\Lambda$CDM. The exponential factor sets the degree of
suppression, while the inflationary input fixes the
dimensional scale of this estimate. The gravitational
contribution of quantum vacuum energy is a separate
question that is not addressed here.

The linear scaling of $\Lambda_{\mathrm{eff}}$ discussed
in Ref.~\cite{rasouli2026emergent} corresponds to the special
case in which the chosen function $f$ has a simple zero at $\alpha=1$.


In addition, the presence of the residual time–dependent correction $\Delta H= (\alpha-1)/(2t)$ in the Hubble parameter suggests that the effective Hubble scale may vary slightly across different cosmological epochs. Such behavior may provide a useful mechanism for discussing the discrepancy between early- and late-time determinations of the Hubble constant, commonly referred to as the Hubble tension.
To estimate the magnitude of this effect, we may use the cosmological bound obtained in our previous analysis for the present universe, namely $|\alpha-1| < 0.2$ \cite{rasouli2026emergent}. As an illustrative example, taking $|\alpha-1|\simeq0.15$ and the age of the universe $t_0\simeq13.8\,{\rm Gyr}$\cite{Planck:2018vyg}, we obtain
$\Delta H \simeq 5\,{\rm km\,s^{-1}\,Mpc^{-1}}$.
This correction is of the same order as the currently observed discrepancy between early- and late-time measurements of the Hubble constant, suggesting that the fractional framework may provide a natural setting for discussing the origin of the Hubble tension.

Taken together, these results indicate that the fractional framework not only unifies gravitational dynamics across widely different scales, but also offers a minimal and physically motivated setting in which cosmological hierarchies and tensions may emerge dynamically from a single underlying parameter.


\section{acknowledgments}
SMMR acknowledges the FCT grant \textbf{UID/212/2025} Centro de Matem\'{a}tica 
e Aplica\c{c}\~{o}es da Universidade da Beira Interior plus
the COST Actions CA23130 (Bridging high and low energies in search of
quantum gravity (BridgeQG)) and CA23115 (Relativistic Quantum Information (RQI)).

\bibliographystyle{spphys}
\bibliography{MinimalRef}
\end{document}